\documentclass[preprint,12pt]{elsarticle}
\usepackage{lineno,hyperref}
\usepackage{amsmath,amssymb}
\usepackage[english]{babel}
\usepackage{bm}
\usepackage{color}

\usepackage{graphicx} 
\usepackage{fancyhdr}
\usepackage{caption}

\usepackage{rotating}
\usepackage{color}

\journal{Chemical Physics Letters}

\usepackage{amsmath,amssymb}
\usepackage[english]{babel}
\usepackage{bm}
\usepackage{color}
\usepackage{indentfirst}

\newcommand{\angstrom}{\mbox{\normalfont\AA}}
 
\usepackage[normalem]{ulem}
\usepackage{color}  

\begin{document}

\begin{frontmatter}

\title{Assessing Parameters for Ring Polymer Molecular Dynamics Simulations at Low Temperatures: DH+H Chemical Reaction}

\author[skoltechaddress]{Ivan S. Novikov\corref{correspondingauthor}}

\cortext[correspondingauthor]{Corresponding author}
\ead{ivan.novikov0590@gmail.com}

\author[yuraaddress]{Yury V. Suleimanov}

\author[skoltechaddress]{Alexander V. Shapeev}

\address[skoltechaddress]{Skolkovo Institute of Science and Technology, Skolkovo Innovation Center, Nobel St. 3, Moscow 143026, Russia}
\address[yuraaddress]{Computation-based Science and
Technology Research Center, Cyprus Institute, 20 Kavafi Street,
Nicosia 2121, Cyprus}

\begin{abstract}

Ring polymer molecular dynamics (RPMD) is an accurate method for calculating thermal chemical reaction rates. It has recently been discovered that low-temperature calculations are strongly affected by the simulation parameters. Here, for the thermally activated reaction DH + H $\to$ D + H$_2$, we calculate the RPMD rate constants at $T = 50$, 100, and 300 K and demonstrate that for $T \geq 100$ K the standard input parameters yield accurate results, but at low temperatures (e.g., 50 K) one must increase the asymptotic distance and force constant, and decrease the umbrella integration step.

\end{abstract}


\end{frontmatter}

\section{Introduction}
Accurate determination of rate constants for gas-phase bimolecular reactions is of great importance in modeling chemical kinetics in various atmospheric, astrochemical, and combustion systems. Chemical reactions that take place at low temperatures are the most challenging case because  quantum mechanical effects such as zero-point energy (ZPE), tunneling, and resonance effects become critically important at such temperatures. Recently, ring polymer molecular dynamics (RPMD) has been proposed as an efficient tool for calculating thermal chemical reaction rates. The method is based on the classical isomorphism between the quantum statistical mechanics of the physical system and the classical statistical mechanics of a fictitious ring polymer including a number of classical copies (beads) of the original system connected by harmonic springs \cite{Wolynes}. RPMD is the full-dimensional approximate quantum mechanical method for computing Kubo-transformed real-time correlation functions for many dynamical properties, in particular, this method computes flux-side correlation functions \cite{miller1983} which are used in the formalism for calculating chemical reaction rates. The path integral nature of the RPMD method allows capturing quantum effects, such as ZPE effects \cite{RPMD:MuH,gonzalez2014theoretical} and tunneling \cite{RPMD:DMuH} that play a key role while calculating thermal rate coefficients at low temperatures. It has been shown that the RPMD technique \cite{Manolopoulos2004,Manolopoulos2005_122,Manolopoulos2005_123,RPMDREVIEW} provides systematically accurate approach for calculating rate coefficients in a wide temperature range. The RPMD method has been implemented in the RPMDrate code \cite{RPMDrate} and tested for numerous prototype chemical reactions \cite{RPMD:MuH,RPMD:DMuH,Suleimanov2017A,gonzalez2014theoretical,
yongle2014quantum,Manolopoulos2011,allen2013communication,
espinosa2020vtst,suleimanov2014XH2,rampino2016thermal,hickson2015ch2,
hickson2017experimentalcd2,hickson2017oh2,bhowmick2018}.

Despite the instantaneous success of the RPMD approach, the calculation of reaction rate constants is computationally demanding at low temperatures due to the specifics of the technique (e.g., a large number of beads is necessary to capture the quantum mechanical effects, long propagation times are needed for the convergence of the rate constants, etc.). The present paper is focused on finding the optimal set of parameters for using the RPMDrate code effectively to predict reaction rate constants at low temperatures down to 50 K. For this aim, we selected the three-atom chemical reaction DH + H $\to$ D + H$_2$ which despite being simple exhibits a number of unexpected effects, such as geometric phase effects \cite{hazra2016gp_effect_h2d,yuan2020gp_effect_h2d} and a narrow Feshbach resonance at energies below the reaction barrier \cite{zhou2020feshbach_resonance_h2d}. This reaction also plays an important role in astrochemistry. The cooling of HD molecule is linked to the gravitational collapse and the fragmentation of clouds \cite{dalgarno1972infrared,lepp2002atomic,mcgreer2008impact}. In the paper \cite{uehara2000dwarf} it was reported that almost all deuterium can be converted to HD molecule and subsequent cooling may lead to the formation of primordial low mass stars and brown dwarfs. 

Note that the system D+H$_2$ has one more channel, namely, the exchange DH + H $\to$ DH + H \cite{aoiz2006cumulative}. However, it wasn't considered in the present study which is focused on achieving the sufficient convergence of the RPMD thermal chemical rate constants for the title channel D + H$_2$.

\section{Ring Polymer Molecular Dynamics}

The ring polymer Hamiltonian for a system of $N$ atoms is written in atomic cartesian coordinates as
\begin{equation} \label{eqn1}
\begin{array}{c}
\displaystyle
H({\bf p},{\bf q}) = \sum_{i=1}^N\sum_{j=1}^{n}\left( \frac{{{\bf p}_i^{(j)}}^2}{2m_i} + \frac{1}{2}m_i\omega^2\left | {\bf q}_i^{(j)} - {\bf q}_i^{(j-1)}\right |^2 \right ) + 
\\
\displaystyle
\sum_{j=1}^{n}V({\bf q}_1^{(j)},\ldots,{\bf q}_N^{(j)}), 
\end{array}
\end{equation}
where $n$ is the number of ring polymer beads, $m_i$ is the mass of the $i$-th atom, ${\bf q}_i^{(j)}$ and ${\bf p}_i^{(j)}$, ~$j=1,\ldots,n$ are the position and the momentum vectors of the $i$-th atom of the system, correspondingly, ${\bf q}_i^{(0)} \equiv {\bf q}_i^{(n)}$, ~$i=1,\ldots,N$. The force constant of the harmonic springs is $\omega={\beta\hbar}/{n}$, where $\beta=1/k_BT$, and $T$ is the temperature of the system. 

Ring polymer molecular dynamics is a classical molecular dynamics in an extended ($n$-bead imaginary time path integral) phase space. The real-time correlation function formalism is used for the RPMD rate coefficient calculation. The expression for the ring polymer flux-side correlation function \cite{miller1983} is given by the $t \to \infty$ limit ($t \to t_{\rm plateau}$ in practice, where $t_{\rm plateau}$ is a ``plateau'' time \cite{chandler1978} when all relevant to the chemical event correlations decay)
\begin{equation} \label{corr_func}
c_{\rm fs}^{(n)}(t) = \dfrac{1}{(2 \pi \hbar)^{fn}} \int d^{fn} {\bf p}_0 \int d^{fn} {\bf q}_0 e^{-(\beta / n) H({\bf p}_0,{\bf q}_0)} \delta[s ({\bf q}_0)] \nu_{s} ({\bf p}_0,{\bf q}_0) h [s ({\bf q}_t)],
\end{equation}
\noindent where $f=3N$, $s({\bf q})$ is the reaction coordinate at ${\bf q}$, $\nu_{s} ({\bf p}_0,{\bf q}_0)$ is the initial velocity along the reaction coordinate
\begin{equation} \label{vel_reaction_coord}
\nu_{s} ({\bf p},{\bf q}) = \sum_{i=1}^N \sum_{j=1}^{n} \dfrac{\partial s({\bf q})}{\partial {\bf q}_i^{(j)}} \dfrac{{\bf p}_i^{(j)}}{m_i},
\end{equation}
and $h [s ({\bf q}_t)]$ is a Heaviside step function, which counts ring polymer trajectories $({\bf p}_t, {\bf q}_t)$ that are on the product side of the dividing surface at time $t$. The RPMD rate coefficient is
\begin{equation} \label{rate_coeff}
k^{(n)}(T) = \dfrac{c_{\rm fs}^{(n)}(t \to \infty)}{Q_r^{(n)}(T)},
\end{equation}
where $Q_r^{(n)}(T)$ is the $n$-bead path integral approximation to the quantum mechanical reactant partition function \cite{Manolopoulos2005_122, Manolopoulos2005_123}. The partition function can be quite difficult to compute accurately \cite{bowman2001importance}. To avoid the direct computation of the partition function we apply the technique implemented in the RPMDrate code \cite{RPMDrate} and briefly described below.

The calculation of RPMD rate coefficient starts with introducing two dividing surfaces. The first dividing surface is located in the asymptotic reactant valley 
\begin{equation} \label{s0}
s_0(\bar{\bf q}) = R_{\infty} - |\bar{\bf R}|,
\end{equation}
where $\bar{\bf q} = (\bar{\bf q}_1, \ldots, \bar{\bf q}_N)$, $\bar{\bf q}_i = \dfrac{1}{n} \sum \limits_{j=1}^{n} {\bf q}_i^{(j)}$, $\bar{\bf R}$ is the centroid of the vector that connects the centres of mass of the reactants, $R_{\infty}$ is an asymptotic distance large enough to make interaction between the reactants negligible. 

The second dividing surface, $s_1(\bar{\bf q})$, is situated near the transition state region and is defined in terms of the bond-breaking and bond-forming distances. In this paper, we consider the reaction with one channel and three atoms A, B, and C: AB + C $\to$ A + BC, i.e., one bond breaks between the atoms A and B and one bond forms between the atoms B and C. Therefore, the second dividing surface is determined as
\begin{equation} \label{s1}
s_1(\bar{\bf q}) = \left( |\bar{\bf q}_{\rm AB}| - q_{\rm AB}^{\ddagger} \right) - \left( |\bar{\bf q}_{\rm BC}| - q_{\rm BC}^{\ddagger} \right),
\end{equation}
where $\bar{\bf q}_{ij}$ is the vector that connects the centroids of $i$-th and $j$-th atoms, $q_{ij}^{\ddagger}$ is the interatomic distance at the transition state.

The next step is the introduction of the reaction coordinate $\xi$ that connects two dividing surfaces
\begin{equation} \label{reaction_coord}
\xi(\bar{\bf q}) = \dfrac{s_0(\bar{\bf q})}{s_0(\bar{\bf q}) - s_1(\bar{\bf q})}.
\end{equation}

As it was mentioned above, the direct computation of the RPMD rate coefficient with \eqref{rate_coeff} can be quite difficult. Therefore, the Bennett-Chandler factorization \cite{bennett1977molecular, chandler1978} is used for the calculation:
\begin{equation} \label{rate_coeff_bennet_chandler}
k_{\rm RPMD}(T) = k_{\operatorname{QTST}}(T;\xi^{\ddagger}) \kappa(t \to t_{\rm plateau};\xi^{\ddagger}).
\end{equation}

The first term, $k_{\operatorname{QTST}}(T;\xi^{\ddagger})$, is the quantum transition state theory rate coefficient (its centroid-density version \cite{gillan87a,gillan87b,voth89}, cd-QTST, when the dividing surfaces are defined using the centroid variables, as in the present case) evaluated near the transition state $\xi^{\ddagger}$.  It is defined in terms of the potential mean force (PMF), or free energy, $W(\xi)$, along the reaction coordinate
\begin{equation} \label{QTST_coeff}
k_{\operatorname{QTST}}(T;\xi^{\ddagger}) = 4\pi R_\infty^2\left (\frac{1}{2\pi\beta\mu_R}\right )^{1/2} e^{-\beta[W(\xi^{\ddagger}) - W(0)]},
\end{equation}
where $\mu_R$ is the reduced mass of the reactants. The PMF difference $W(\xi^{\ddagger}) - W(0)$ is calculated using umbrella integration \cite{Kastner2005,Kastner2006,Kastner2009}
\begin{equation} \label{PMF_difference}
W(\xi^{\ddagger}) - W(0) = \int \limits_0^{\xi^{\ddagger}} \sum \limits_{i=1}^{N_{\rm windows}} \left [ \dfrac{n_i \mathcal{N}({\bar{\xi}}_i, \sigma_i^2)}{\sum \limits_{i=1}^{N_{\rm windows}} n_j \mathcal{N}({\bar{\xi}}_j, \sigma_j^2)} \left( \dfrac{1}{\beta} \dfrac{\xi - {\bar{\xi}}_i}{\sigma_i^2} - k_i (\xi_i - {\bar{\xi}}_i) \right) \right ] d \xi,
\end{equation}
where $N_{\rm windows}$ is the number of biasing windows, $n_i$ is the total number of steps sampled for the $i$-th window, $\mathcal{N}({\bar{\xi}}_i, \sigma_i^2)$ is the normal distribution of the reaction coordinate $\xi$ with the mean value (or, center of the biasing window) ${\bar{\xi}}_i$ and the variance $\sigma_i^2$, and $k_i$ is the force constant which defines the strength of the bias.

The second term in \eqref{rate_coeff_bennet_chandler}, $\kappa(t \to t_{\rm plateau};\xi^{\ddagger})$, is the ring polymer recrossing factor. This factor plays a role of a dynamical correction to cd-QTST and ensures that the resulting RPMD rate coefficient $k_{\rm RPMD}(T)$ will be independent of the choice of the dividing surface \cite{Manolopoulos2004}. It is expressed as the ratio between two real-time flux-side correlation functions
\begin{equation} \label{transmission_coeff}
\kappa(t \to t_{\rm plateau};\xi^{\ddagger}) = \dfrac{c_{\rm fs}^{(n)}(t \to t_{\rm plateau}; \xi^{\ddagger})}{c_{\rm fs}^{(n)}(t \to 0_+; \xi^{\ddagger})}.
\end{equation}

Thus, after calculating the cd-QTST rate coefficient and the recrossing factor, we can calculate the RPMD rate coefficient.

\section{Steps of the RPMDrate code calculation and parameters to be optimized}

The RPMDrate code works in five sequential steps:

\begin{itemize}

\item[1.] Generating the initial configurations for umbrella integration;

\item[2.] Umbrella integration along the reaction coordinate; 

\item[3.] Calculation of the cd-QTST rate coefficient;

\item[4.] Computation of the recrossing factor;

\item[5.] Calculation of the resulting RPMD thermal rate coefficient.

\end{itemize}
We describe all the RPMDrate stages below and the crucial code parameters that significantly affect the resulting RPMD rate constant and, thus, should be optimized.

We start from the introduction of the umbrella sampling interval with the left end $\xi_{\rm min}$ and the right end $\xi_{\rm max}$ and divide it into a series of windows $\xi_i$, $i=1,\ldots,N_{\rm windows}$. At the first RPMDrate step we generate classical configurations ($n=1$) for each window at $T = 300$ K. These configurations are used in subsequent umbrella integration RPMD simulations. 

At the second RPMDrate step, we equilibrate the system for each $i$-th window (during the $t_{\rm equilibration}$ time) and accumulate ${\bar{\xi}}_i$ and $\sigma_i$ along the RPMD thermostatted trajectory run for $t_{\rm sampling}$. This process is repeated $N_{\rm trajectory}$ times for accumulating enough steps of sampling and is the main part of umbrella integration \cite{Kastner2005,Kastner2006,Kastner2009}. For the reaction investigated here, the trajectory is thermostatted using one of the generalized Langevin equation (GLE) thermostats \cite{ceriotti2009langevin,ceriotti2009colored,ceriotti2010colored}. As the result of the second RPMDrate stage, we obtain the potential of mean force, or, free energy, $W(\xi)$, and its maximum, $W(\xi^{\ddagger})$, which depends on temperature $T$. The cd-QTST rate coefficient, calculated at the next RPMDrate stage, exponentially depends on the difference between the maximum $\xi = \xi^{\ddagger}$ of free energy and its value at $\xi = 0$ (see \eqref{QTST_coeff}, \eqref{PMF_difference}) and, thus, the difference $W(\xi^{\ddagger}) - W(0)$ is the crucial value for estimation of the cd-QTST rate coefficient. As it was demonstrated in the previous papers \cite{RPMDrate,suleimanov2014XH2,hickson2015ch2,hickson2017experimentalcd2,hickson2017oh2}, the accuracy of the mentioned difference calculation depends on the number of beads $n$, especially at low temperatures. Moreover, the choice of the asymptotic distance $R_{\infty}$ and, therefore, the umbrella integration step $d \xi$ and the force constant $k_i$ also affects the resulting cd-QTST rate coefficient \cite{bhowmick2018}. Thus, the parameters $n$, $R_{\infty}$, $d \xi$, and $k_i$ are the most important ones to be optimized during the first three RPMDrate steps. 

Next, at the fourth RPMDrate step, we calculate the recrossing factor. For this calculation, we use a constrained RPMD simulation in the presence of a thermostat (parent trajectory) to generate a series of independent configurations at the initial time moment ${\bf q}_0$ with centroids on the transition state dividing surfaces $s_{\xi^{\ddagger}}(\bar{\bf q}) = \xi^{\ddagger} s_1(\bar{\bf q}) + (1 - \xi^{\ddagger}) s_0(\bar{\bf q}) = 0$. The RATTLE algorithm \cite{andersen1983rattle} is introduced into the time integration to constrain the centroid $\bar{{\bf q}}_0$ to the dividing surfaces $s_{\xi^{\ddagger}}(\bar{\bf q}) = 0$. For each of these constrained configurations obtained from the parent trajectory, a number of momentum vectors ${\bf p}_0$ is randomly sampled from the Maxwell distribution and the resulting recrossing (child) trajectories are evolved forward in the propagation (``plateau'') time $t_{\rm plateau}$ without the thermostat or the dividing surface constraint. One of the reasons of inaccurate determining the recrossing factor may be a short ``plateau'' time, because the decay of the recrossing factor can occur rather slowly, and it is important to guarantee that the recrossing factor is a constant during a long time interval. Thus, the time $t_{\rm plateau}$ is one of the most important parameters to be optimized at the fourth RPMDrate step.

Finally, at the fifth step we calculate the resulting RPMD thermal rate coefficient. Note that the strategy is also suitable for the classical dynamics (with $n = 1$ beads) and at the classical high-temperature limit ($T \to \infty$).

The RPMDrate code was successfully applied to many chemical reactions. The parameters 
used for calculation of different chemical reaction rates are shown in Table 
\ref{tabs1} \cite{RPMD:MuH,RPMD:DMuH,Suleimanov2017A,gonzalez2014theoretical,
yongle2014quantum,Manolopoulos2011,allen2013communication,
espinosa2020vtst,suleimanov2014XH2,rampino2016thermal,hickson2015ch2,
hickson2017experimentalcd2,hickson2017oh2,bhowmick2018}. Thermally activated reactions were mainly considered at the temperatures $T \geq 100$ K. For most of these reactions the standard number of beads, umbrella integration step, and force constant were chosen: $n=128$, $d \xi = 0.01$, $k_i = 2.72$ ($T/$K) eV. Thus, the above umbrella integration step and the force constant could be chosen for calculations with any asymptotic distance $R_{\infty}$ at the temperatures $T = 100$ and 300 K. As opposed to the thermally activated reactions, barrirless reactions were also considered at low temperatures. In the paper Ref. \cite{bhowmick2018} it was shown that while increasing the asymptotic distance one should also increase the force constant for low temperatures. Moreover, if the temperature of the reaction under investigation is decreased, one should increase the number of beads for obtaining the correct rate constants. Finally, the longer ``plateau'' time is needed for barrierless reactions as compared to thermally activated ones due to slower decay of the oscillations in the ring polymer recrossing factor.

\begin{table}[h!]
\caption{\label{tabs1} Input RPMDrate code parameters for thermally activated and barrierless reactions. The force constant $k_i$ is reported in ($T/$K) eV, the ``plateau'' time $t_{\rm plateau}$  is reported in ps, and the temperature $T$ is reported in K. } 
\begin{center}
\begin{tabular}{c|c|c|c|c|c} \hline \hline
\footnotesize{Thermally activated reaction} & \footnotesize{$R_{\infty}$} & \footnotesize{$d \xi$} & \footnotesize{$k_i$} & \footnotesize{$n$($T$)} & \footnotesize{$t_{\rm plateau}$} \\ \hline
\footnotesize{Mu+H$_2 \to $ MuH+H \cite{RPMD:MuH}} & \footnotesize{$30 a_0$} & \footnotesize{0.01} & \footnotesize{2.72} & \footnotesize{512(200-300)} & \footnotesize{0.06} \\ \hline
\footnotesize{D+HMu $\to$ DMu+H \cite{RPMD:DMuH}} & \footnotesize{$30 a_0$} & \footnotesize{0.01} & \footnotesize{2.72} & \footnotesize{128(150-300)} & \footnotesize{0.05} \\ \hline
\footnotesize{OH+H$_2 \to $ H+H$_2$O \cite{Suleimanov2017A}} & \footnotesize{$11 a_0$} & \footnotesize{0.01} & \footnotesize{2.72} & \footnotesize{128(150-300)} & \footnotesize{0.05} \\ \hline
\footnotesize{O($^3$P)+CX$_4$ $\to$ OX+CX$_3$ (X=H,D) \cite{gonzalez2014theoretical}} & \footnotesize{$15 a_0$} & \footnotesize{0.01} & \footnotesize{2.72} & \footnotesize{128(200-300)} & \footnotesize{0.1} \\ \hline
\footnotesize{Cl+CX$_4 \to $ XCl+CX$_3$ (X=H,D) \cite{yongle2014quantum}} & \footnotesize{$30 a_0$} & \footnotesize{0.01} & \footnotesize{2.72} & \footnotesize{64(300)} & \footnotesize{0.1} \\ \hline
\footnotesize{H+CH$_4 \to $ H$_2$+CH$_3$ \cite{Manolopoulos2011}} & \footnotesize{$30 a_0$} & \footnotesize{0.01} & \footnotesize{2.72} & \footnotesize{128(200-300)} & \footnotesize{0.1} \\ \hline
\footnotesize{OH+CX$_4 \to $ HXO+CX$_3$ (X=H,D) \cite{allen2013communication}} & \footnotesize{$15 a_0$} & \footnotesize{0.01} & \footnotesize{2.72} & \footnotesize{128(200-300)} & \footnotesize{0.05} \\ \hline
\footnotesize{X+C$_2$H$_6$ $\to$ HX + C$_2$H$_5$ (X=H,Cl,F) \cite{espinosa2020vtst}} & \footnotesize{$15 a_0$} & \footnotesize{0.01} & \footnotesize{2.72} & \footnotesize{128(200-300)} & \footnotesize{0.05 (H)} \\
 & & & & & \footnotesize{0.1 (Cl)} \\ 
  & & & & & \footnotesize{0.2 (F)} \\ \hline\hline
\footnotesize{Barrierless reaction} & \footnotesize{$R_{\infty}$} & \footnotesize{$d \xi$} & \footnotesize{$k_i$} & \footnotesize{$n$($T$)} & \footnotesize{$t_{\rm plateau}$} \\ \hline
\footnotesize{X($^1$D)+H$_2 \to $ XH+H (X=C,S) \cite{suleimanov2014XH2}} & \footnotesize{$20 a_0$} & \footnotesize{0.01}  & \footnotesize{2.72} & \footnotesize{256(50); 128(300)} & \footnotesize{2} \\ \hline
\footnotesize{C+CH$^+ \to$ C$_2^+$+H \cite{rampino2016thermal}} & \footnotesize{$15 a_0$} & \footnotesize{0.01}  & \footnotesize{2.72} & \footnotesize{128(20-300)} & \footnotesize{16} \\ \hline
\footnotesize{C($^1$D)+H$_2 \to $ CH+H \cite{hickson2015ch2}} & \footnotesize{$20 a_0$} & \footnotesize{0.01}  & \footnotesize{2.72} & \footnotesize{256(50); 128(300)} & \footnotesize{2} \\ \hline
\footnotesize{C($^1$D)+D$_2 \to $ CD+D \cite{hickson2017experimentalcd2}} & \footnotesize{$15 a_0$} & \footnotesize{0.01}  & \footnotesize{2.72} & \footnotesize{128(50); 48(300)} & \footnotesize{2} \\ \hline
\footnotesize{O($^1$D)+X$_2 \to $ OX+X (X=H,D) \cite{hickson2017oh2}} & \footnotesize{$15 a_0$} & \footnotesize{0.01}  & \footnotesize{2.72} & \footnotesize{296(50); 128(300)} & \footnotesize{2} \\ \hline
\footnotesize{D$^+$+H$_2 \to$ HD+H$^+$ \cite{bhowmick2018}} & \footnotesize{12-30 $\angstrom$} & \footnotesize{0.01}  & \footnotesize{2.72-20.4} & \footnotesize{160(20,50); 128(75,100)} & \footnotesize{3-8} \\ \hline \hline

\end{tabular}
\end{center}
\end{table}

In this paper we investigate how the parameters $n$, $R_{\infty}$, $d \xi$, $k_i$, and $t_{\rm plateau}$ affect the resulting RPMD rate coefficient at different temperatures for the thermally activated reaction DH+H $\to$ D+H$_2$.

\section{Results and discussion}

We used the BKMP2 \cite{boothroyd1996h3} potential energy surface for all the computations described here. First, we optimize the parameters $R_{\infty}$, $d \xi$, $k_i$ at the fixed temperatures ($T = 50, 100$, and 300 K) and $n = 64$ (we consider this number of beads as it is typically enough for approximate estimation the cd-QTST rate coefficient). The optimized parameters at these temperatures are summarized in Table \ref{tabs2}. As for the reactions in Table \ref{tabs1}, at the temperatures $T \geq 100$ K the standard parameters $d \xi = 0.01$ and $k_i = 2.72$ ($T/$K) eV are suitable for computations with $10a_0 \leq R_{\infty} \leq 30 a_0$, the resulting $k_{\rm QTST}$ coefficients are similar for all temperatures of interest. However, at low temperature $T = 50$ K, the force constant increases and the umbrella integration step decreases while the asymptotic distance is increased. Thus, at low temperatures, more attention should be paid to the choice of  $R_{\infty}$, $d \xi$, and $k_i$. We demonstrate further results for the standard parameters used in the past ($R_{\infty} = 15 a_0$, $d \xi = 0.01$, $k_i = 2.72$ ($T/$K) eV) at $T = 50, 100$, and 300 K, and for the modified parameters ($R_{\infty} = 30 a_0$, $d \xi = 0.005$, $k_i = 13.6$ ($T/$K) eV) at $T = 50$ K. We note, that $R_{\infty} \leq 30 a_0$ is sufficient for obtaining accurate enough $k_{\rm QTST}$ at the temperature $T = 50$ K, therefore, we provide the resulting RPMD rate coefficients only for $R_{\infty} \leq 30 a_0$.

\begin{table}[h!]
\caption{\label{tabs2} Asymptotic distance $R_{\infty}$, umbrella integration step $d \xi$, and force constant $k_i$ optimized at $T = 50, 100$, and 300 K and $n = 64$. The force constants are reported in ($T/$K) eV, the cd-QTST rate coefficients are reported in cm$^3$ s$^{-1}$. The standard umbrella integration step $d \xi = 0.01$ and the force constant $k_i = 2.72$ ($T/$K) eV could be chosen for calculation with any asymptotic distance at the temperatures $T \geq 100$ K. At the low temperature, $T = 50$ K, the parameter $k_i$ increases and the step $d \xi$ decreases while increasing the distance $R_{\infty}$ for obtaining the similar cd-QTST rate coefficients. The resulting cd-QTST rate coefficient oscillates near $1.44 \times 10^{-26}$ cm$^3$ s$^{-1}$, thus, the asymptotic distance $R_{\infty} = 30 a_0$ is sufficient for obtaining accurate enough rate coefficient at $T = 50$ K.} 
\begin{center}
\begin{tabular}{c|c|c|c|c|c|c|c} \hline \hline 
\multicolumn{4}{c|}{\footnotesize{$T = 50$ K}} & \multicolumn{2}{c|}{\footnotesize{$T = 100$ K, $d \xi = 0.01$, $k_i = 2.72$}} & \multicolumn{2}{c}{\footnotesize{$T = 300$ K, $d \xi = 0.01$, $k_i = 2.72$}} \\ \hline
\footnotesize{$R_{\infty}$} & \footnotesize{$d \xi$} & \footnotesize{$k_i$} & \footnotesize{$k_{\rm QTST}$} & \footnotesize{$R_{\infty}$} &  \footnotesize{$k_{\rm QTST}$} & \footnotesize{$R_{\infty}$} & \footnotesize{$k_{\rm QTST}$} \\ \hline
\footnotesize{$15 a_0$} & \footnotesize{0.01} & \footnotesize{2.72} & \footnotesize{$1.34 \times 10^{-26}$}  & \footnotesize{$15 a_0$} & \footnotesize{$1.19 \times 10^{-23}$} & \footnotesize{$15 a_0$} & \footnotesize{$5.06 \times 10^{-17}$} \\
\footnotesize{$20 a_0$} & \footnotesize{0.01} & \footnotesize{10.88} & \footnotesize{$1.44 \times 10^{-26}$} & \footnotesize{$20 a_0$} & \footnotesize{$1.24 \times 10^{-23}$} & \footnotesize{$20 a_0$} & \footnotesize{$5.05 \times 10^{-17}$} \\  
\footnotesize{$30 a_0$} & \footnotesize{0.005} & \footnotesize{13.60} & \footnotesize{$1.46 \times 10^{-26}$} & \footnotesize{$30 a_0$} & \footnotesize{$1.22 \times 10^{-23}$} & \footnotesize{$30 a_0$} & \footnotesize{$5.09 \times 10^{-17}$} \\
\footnotesize{$40 a_0$} & \footnotesize{0.005} & \footnotesize{29.92} & \footnotesize{$1.42 \times 10^{-26}$} & & & & \\ \hline \hline 
\end{tabular} 
\end{center}
\end{table}

\begin{figure}[h!] \begin{center}
\includegraphics[width=2.3in, height=2.5in, keepaspectratio=false, angle=270]{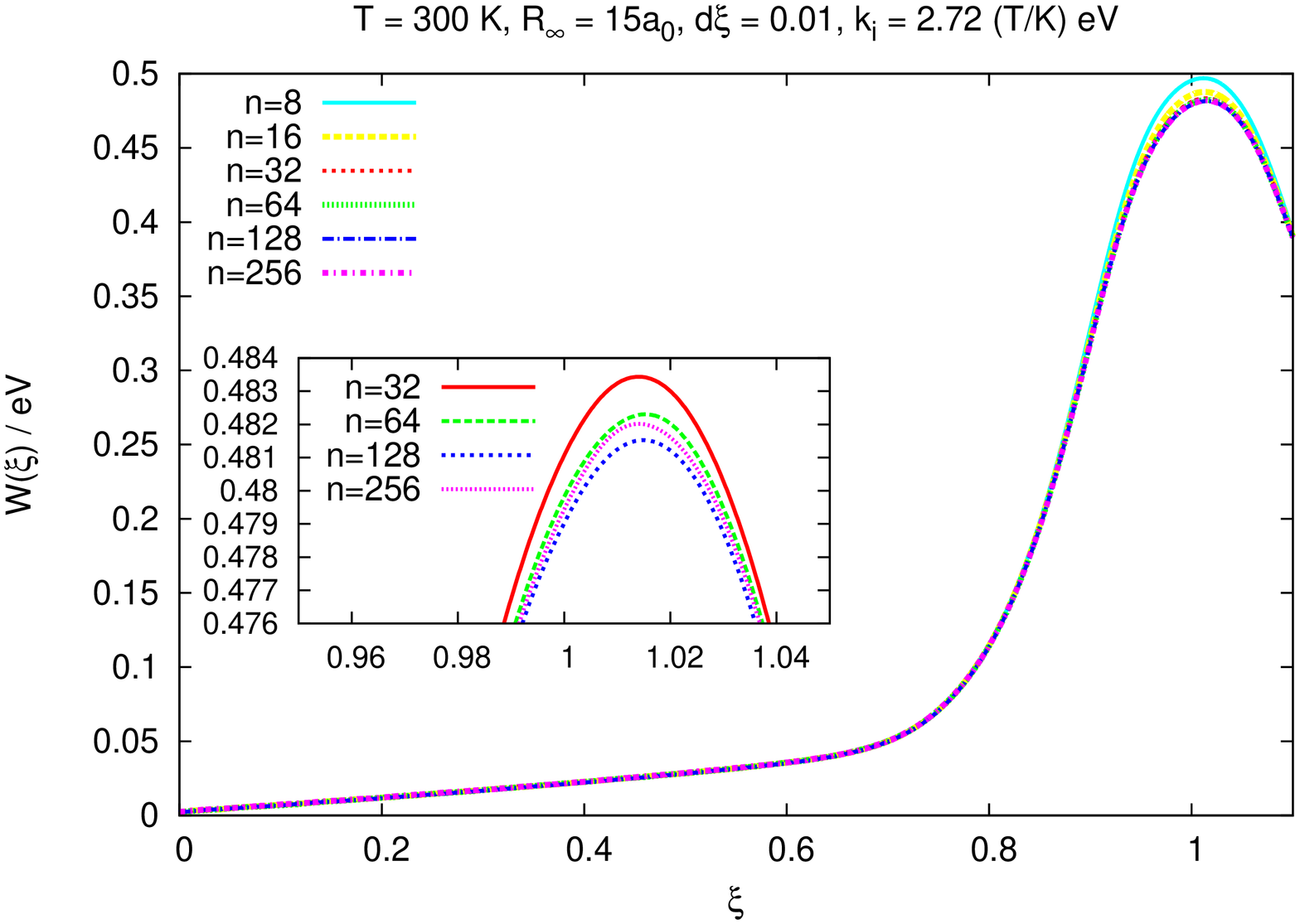}
\includegraphics[width=2.3in, height=2.5in, keepaspectratio=false, angle=270]{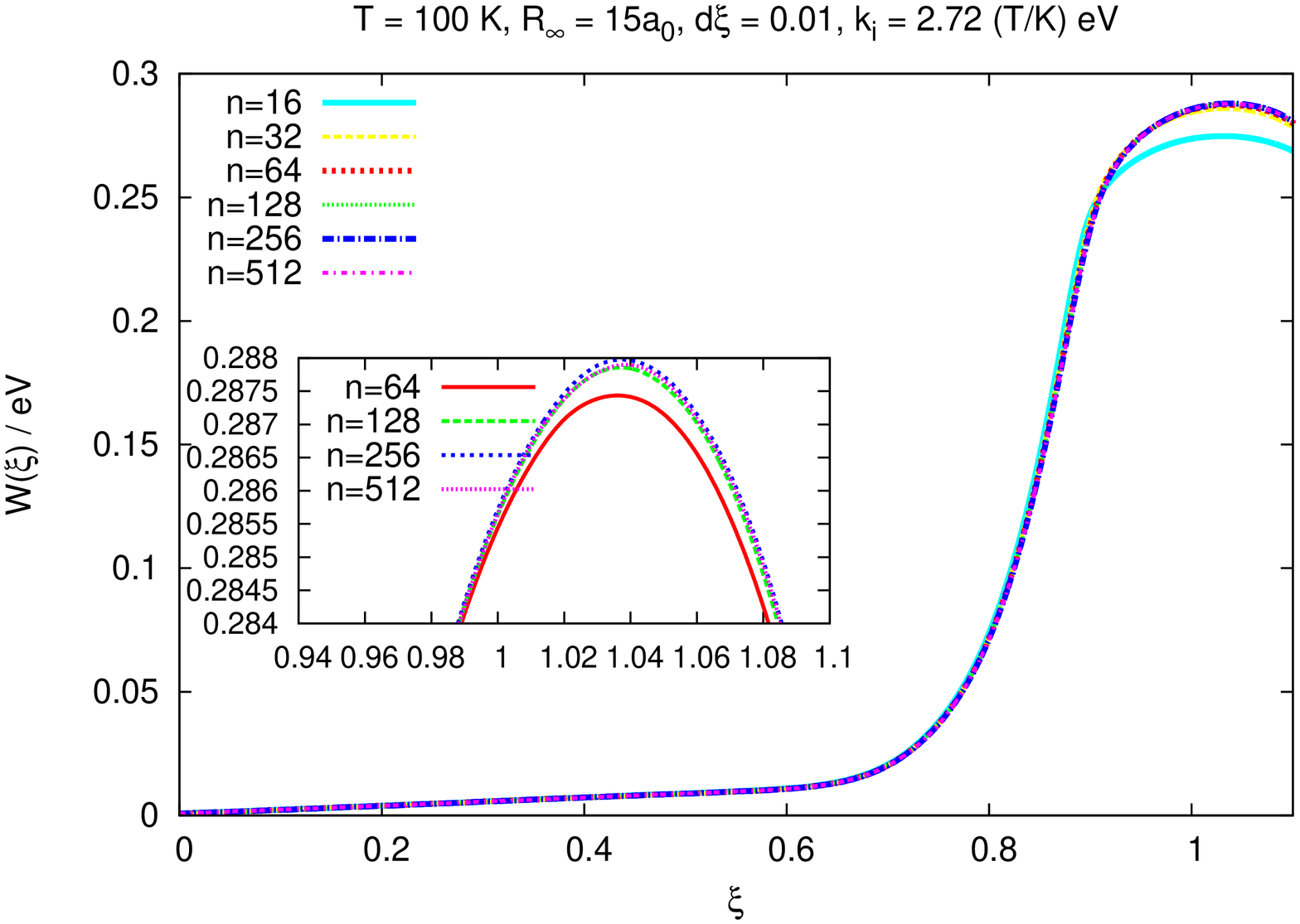}
\includegraphics[width=2.3in, height=2.5in, keepaspectratio=false, angle=270]{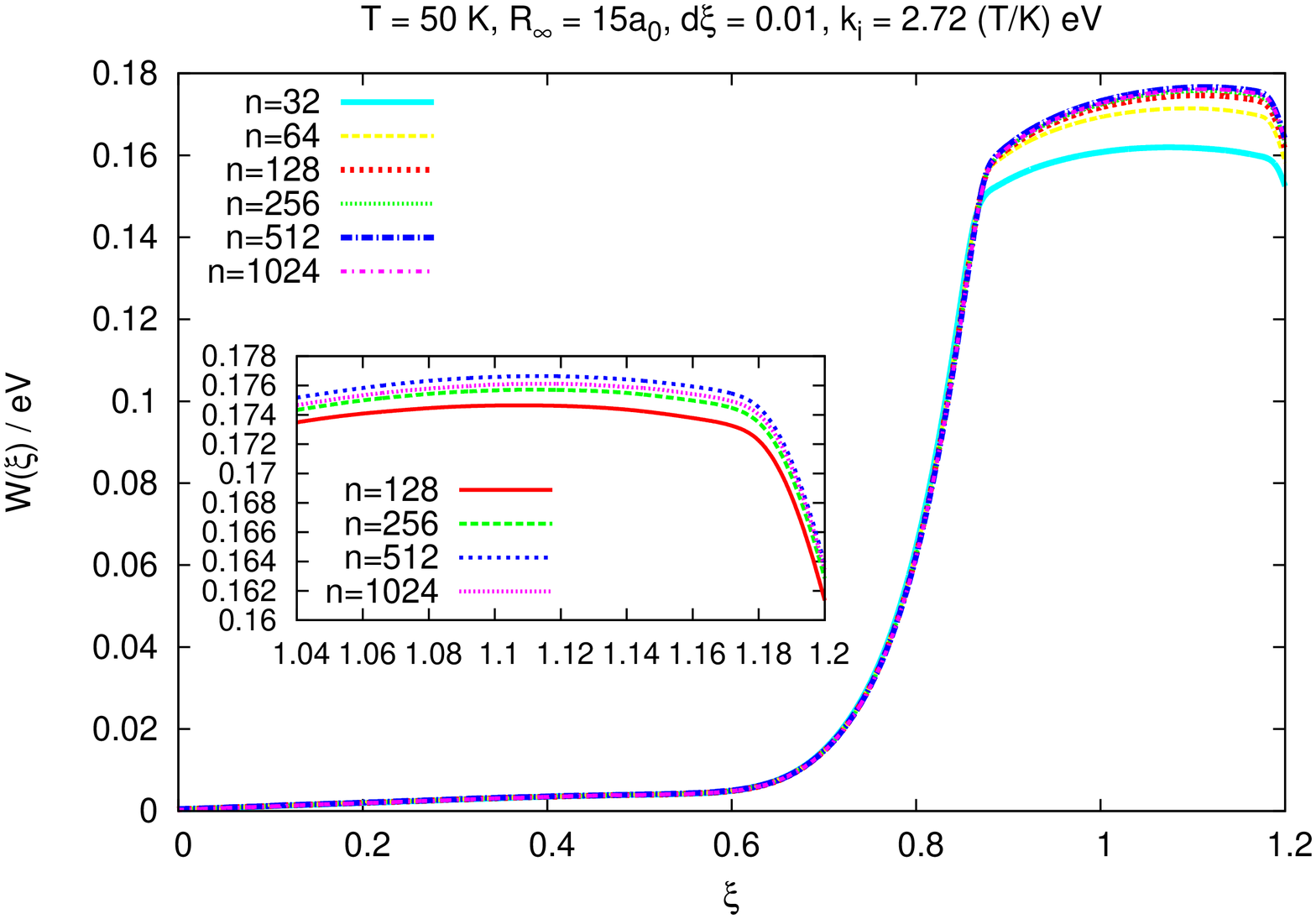}
\includegraphics[width=2.3in, height=2.5in, keepaspectratio=false, angle=270]{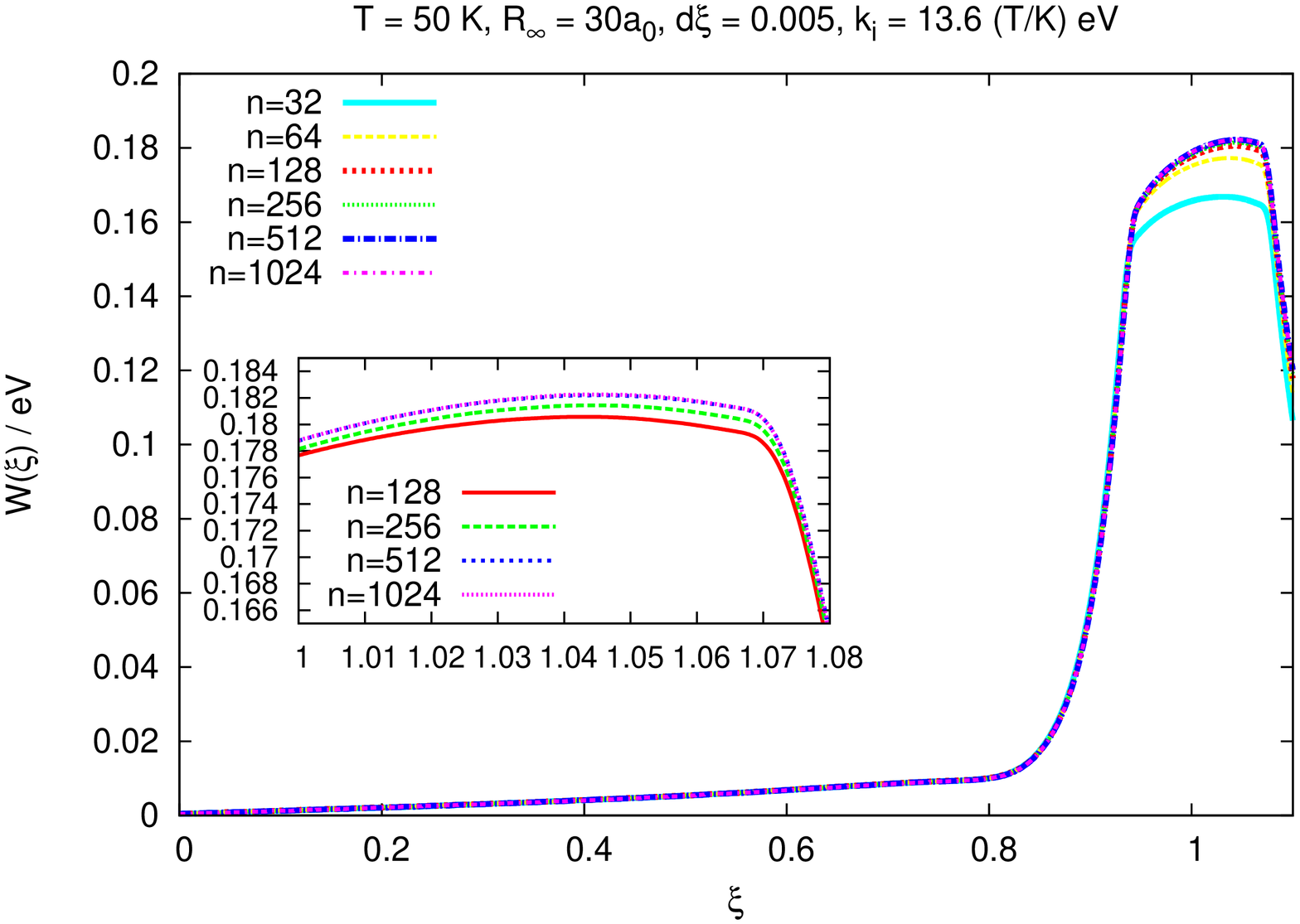}
\caption{\label{fig:bead_convergence} Free energy profiles for different number of beads, standard parameters $R_{\infty} = 15a_0$, $d \xi = 0.01$, and $k_i = 2.72$ ($T/$K) eV ($T = 50, 100$, and 300 K), and modified parameters $R_{\infty} = 30a_0$, $d \xi = 0.005$, and $k_i = 13.6$ ($T/$K) eV ($T = 50$ K). The energy barrier increases while increasing the temperature. The convergence of the energy profiles with respect to the number of beads is observed for $n = 128$ at $T = 300$ K, $n = 256$ at $T = 100$ K, and $n = 512$ at $T = 50$ K. The converged cd-QTST rate coefficients are similar for two various sets of parameters at $T = 50$ K: $k_{\rm QTST} = 4.16 \times 10^{-27}$ cm$^3$ s$^{-1}$ for $R_{\infty} = 15a_0$, $k_{\rm QTST} = 4.52 \times 10^{-27}$ cm$^3$ s$^{-1}$ for $R_{\infty} = 30a_0$ ($n = 512$).}
\end{center} \end{figure}

Next, we investigate the convergence of free energy profiles used to calculate the cd-QTST rate coefficient with respect to the number of beads. The convergence is illustrated in Fig. \ref{fig:bead_convergence}. From the plots we conclude:
while decreasing the temperatures one should increase the number of beads for the proper estimation of the difference $W(\xi^{\ddagger}) - W(0)$, and, therefore, the cd-QTST rate coefficient. Here we choose $n = 128$ at $T = 300$ K, $n = 256$ at $T = 100$ K, and $n = 512$ at $T = 50$ K. We also observe a conventional trend: free energy barrier increases as the temperature increases.

\begin{figure}[h!] \begin{center}
\includegraphics[width=2.3in, height=2.5in, keepaspectratio=false, angle=270]{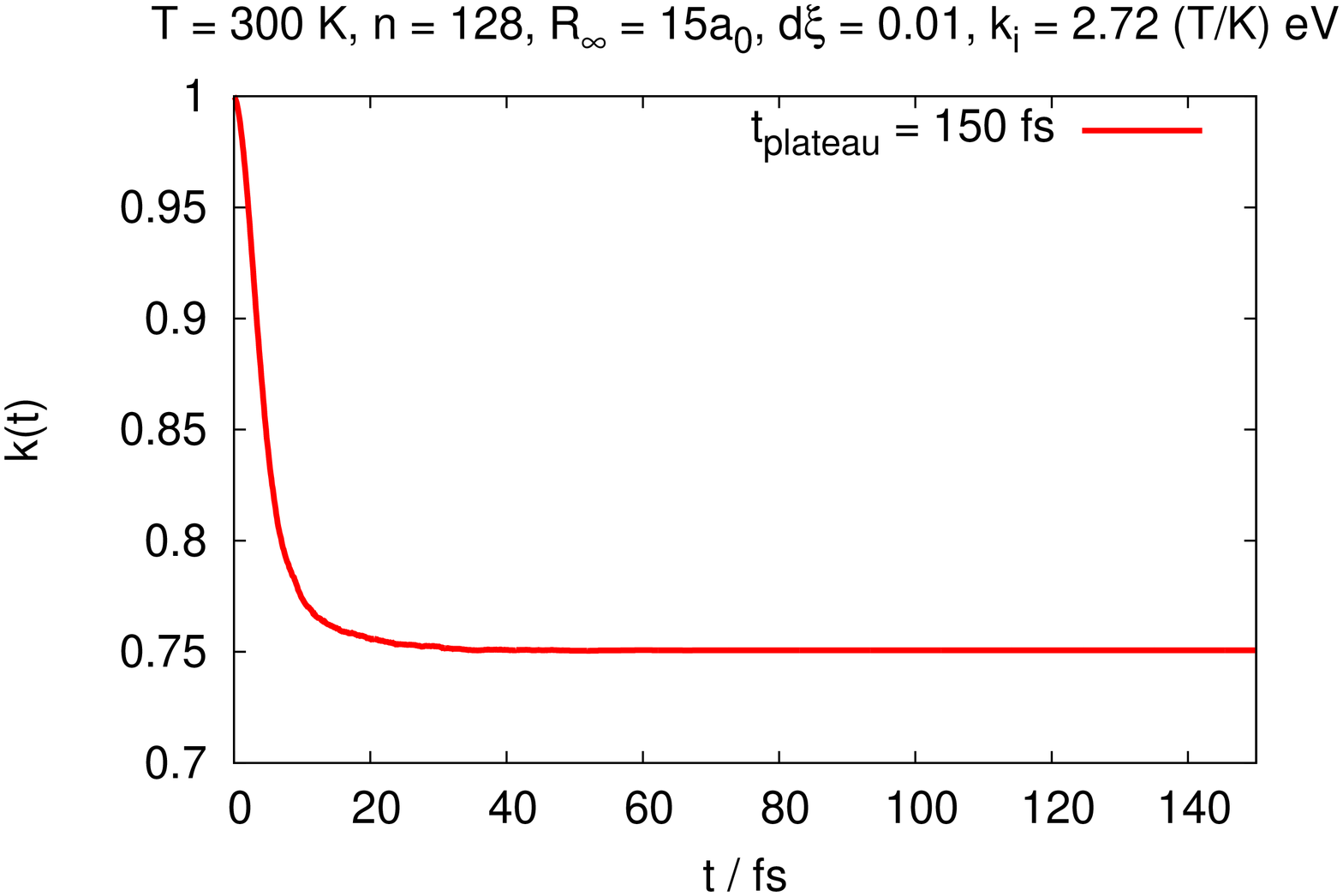}
\includegraphics[width=2.3in, height=2.5in, keepaspectratio=false, angle=270]{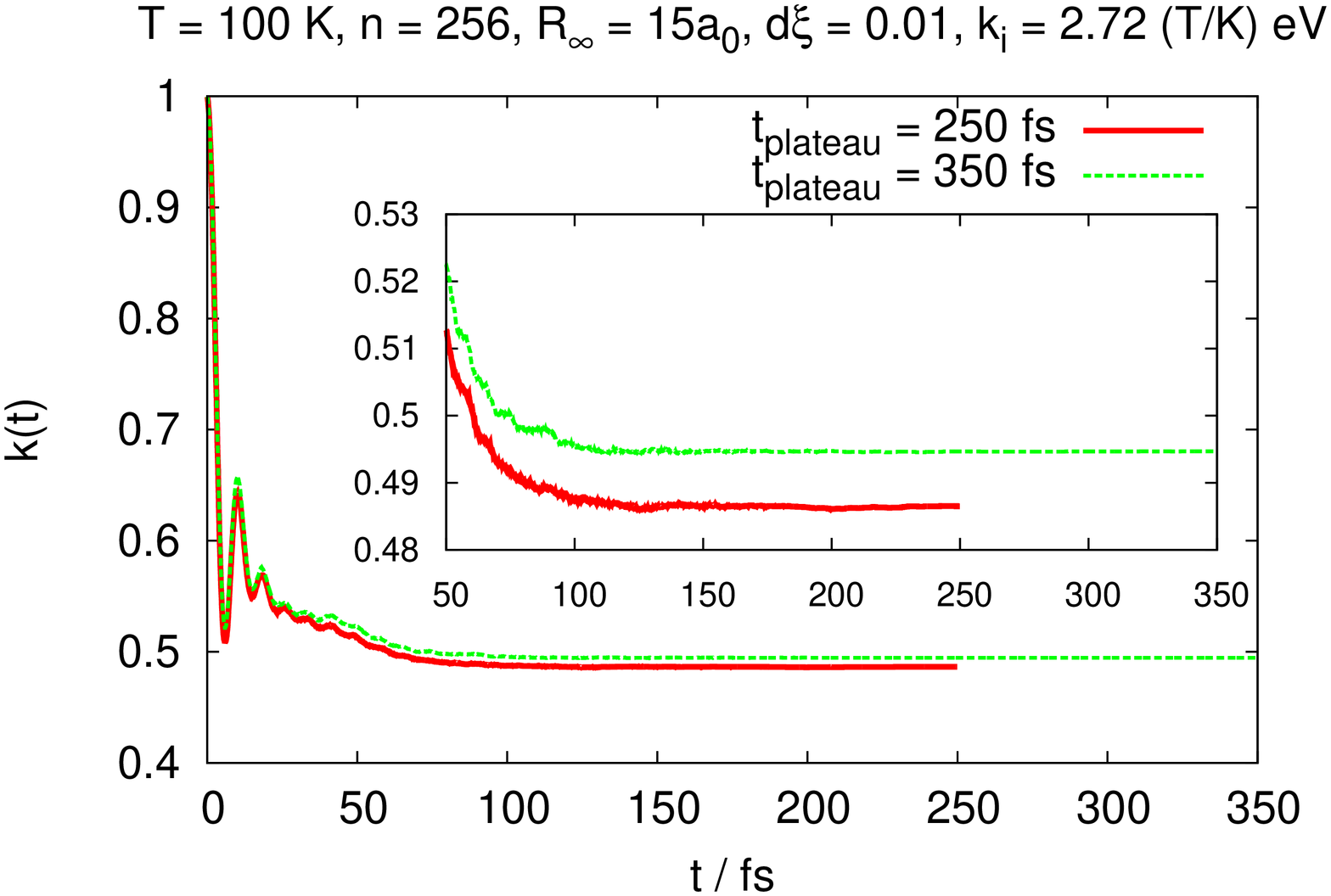}
\includegraphics[width=2.3in, height=2.5in, keepaspectratio=false, angle=270]{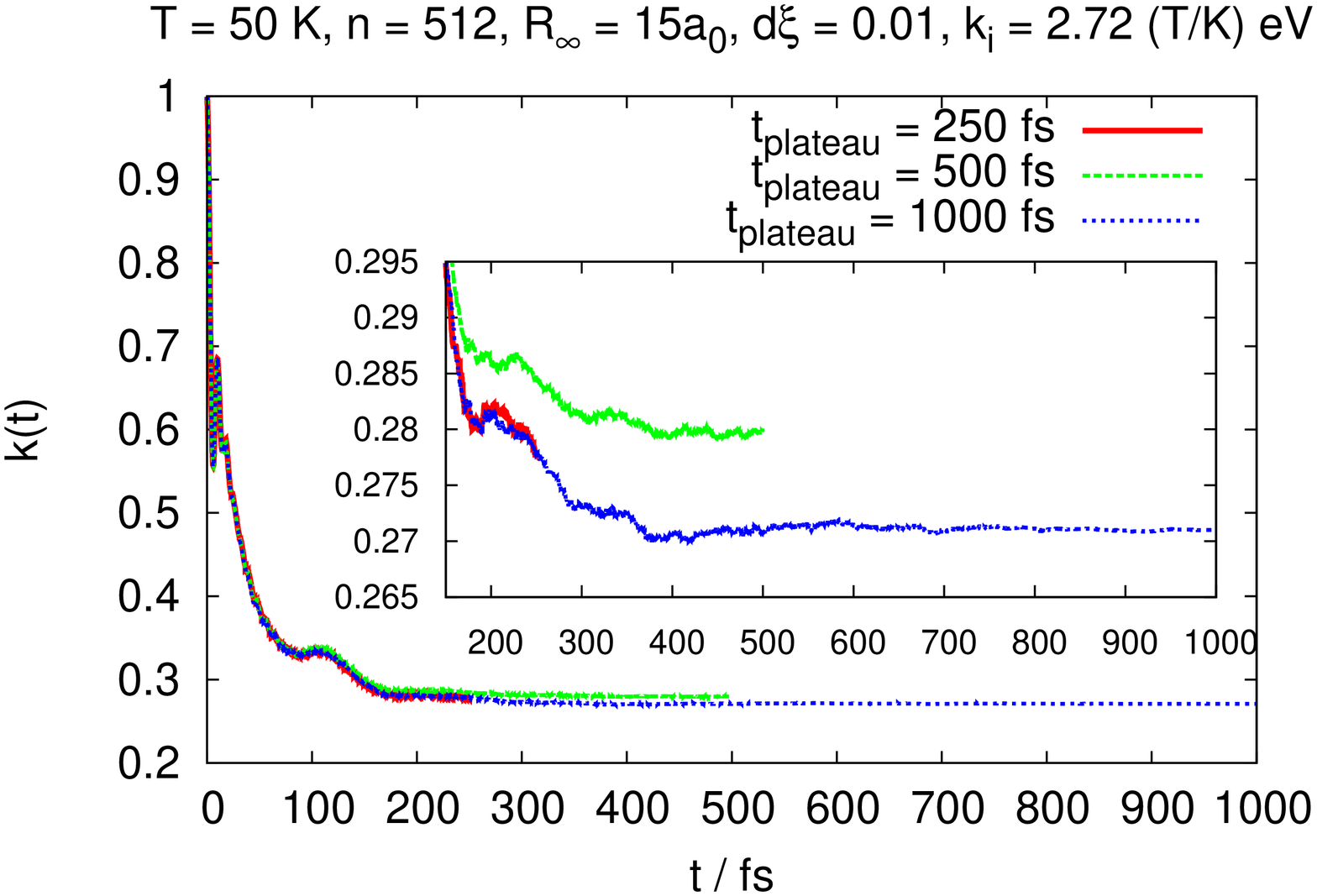}
\includegraphics[width=2.3in, height=2.5in, keepaspectratio=false, angle=270]{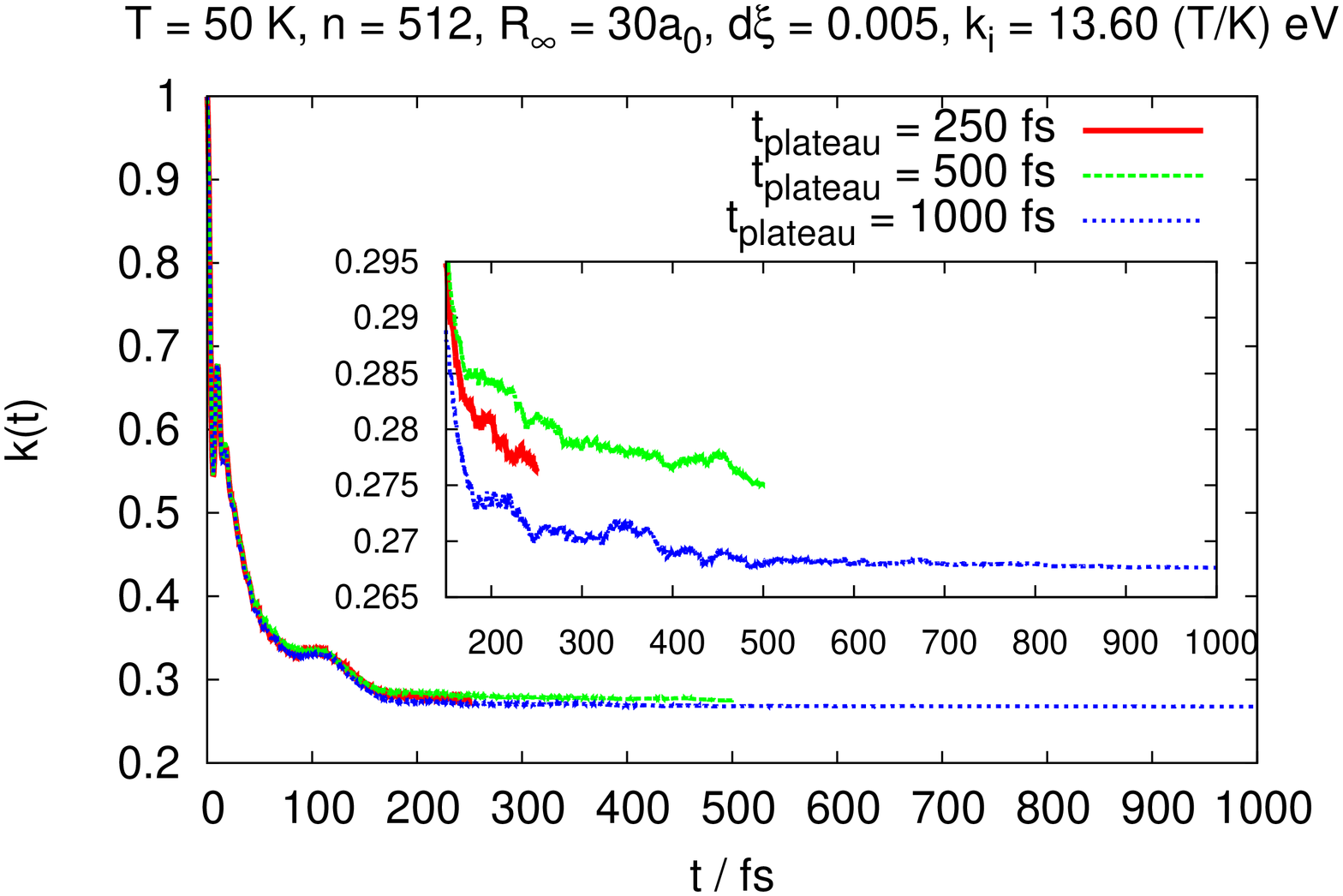}
\caption{\label{fig:plateau_time_convergence} Ring polymer recrossing factors for different different ``plateau'' times, standard parameters $R_{\infty} = 15a_0$, $d \xi = 0.01$, and $k_i = 2.72$ ($T/$K) eV ($T = 50, 100$, and 300 K), and modified parameters $R_{\infty} = 30a_0$, $d \xi = 0.005$, and $k_i = 13.6$ ($T/$K) eV ($T = 50$ K). When the temperature of the reaction is decreased, one should increase the ``plateau'' time for obtaining the correct recrossing factor. The resulting ring polymer recrossing factor is increased if the temperature is also increased.}
\end{center} \end{figure}

Finally, we study the convergence of the ring polymer recrossing factors with respect to the ``plateau'' time at different temperatures. The recrossing factors in the investigated temperature range are shown in Fig. \ref{fig:plateau_time_convergence}. The longer ``plateau'' time is needed for obtaining the correct ring polymer recrossing factor if the temperature is decreased: $t_{\rm plateau} \approx 0.15$ ps at $T = 300$ K, $t_{\rm plateau} \approx 0.35$ ps at $T = 100$ K, and $t_{\rm plateau} \approx 1$ ps at $T = 50$ K. 

The resulting sets of input parameters for all the RPMDrate steps, cd-QTST rate coefficients, recrossing factors, and RPMD rate coefficients at $T = 50, 100$, and 300 K are summarized in Table \ref{tabs4}. As it was expected for the thermally activated reaction, the RPMD rate coefficient increases as the temperature increases. We also note that the resulting $k_{\rm RPMD}$ rates are similar at $T = 50$ K for both the standard parameters $R_{\infty}$, $d \xi$, $k_i$ and the modified ones. However, the relative deviation between them is $\approx 10 \%$. Therefore, if a very high accuracy of calculating the RPMD rate coefficient is required one should increase the asymptotic distance at low temperatures. 

\begin{table}[h!]
\caption{\label{tabs4} Input RPMDrate parameters, resulting recrossing factors and rate coefficients at $T = 50, 100$, and 300 K.} 
\begin{center}
\begin{tabular}{c|c|c|c|c|c|c|c|c} \hline \hline 
\footnotesize{$T$, K} & \footnotesize{$n$} & \footnotesize{$R_{\infty}$} & \footnotesize{$d \xi$} & \footnotesize{$k_i$, ($T/$K) eV} & \footnotesize{$t_{\rm plateau}$, ps} & \footnotesize{$k_{\rm QTST}$, cm$^3$ s$^{-1}$} & \footnotesize{$\kappa$} & \footnotesize{$k_{\rm RPMD}$, cm$^3$ s$^{-1}$} \\ \hline
\footnotesize{300} & \footnotesize{128} & \footnotesize{$15 a_0$} & \footnotesize{0.01} & \footnotesize{2.72} & \footnotesize{0.15} & \footnotesize{$5.18 \times 10^{-17}$}  & \footnotesize{0.751} & \footnotesize{$3.89 \times 10^{-17}$} \\
\footnotesize{300} & \footnotesize{128} & \footnotesize{$20 a_0$} & \footnotesize{0.01} & \footnotesize{2.72} & \footnotesize{0.15} & \footnotesize{$5.19 \times 10^{-17}$}  & \footnotesize{0.751} & \footnotesize{$3.90 \times 10^{-17}$} \\
\footnotesize{300} & \footnotesize{128} & \footnotesize{$30 a_0$} & \footnotesize{0.01} & \footnotesize{2.72} & \footnotesize{0.15} & \footnotesize{$5.35 \times 10^{-17}$}  & \footnotesize{0.749} & \footnotesize{$4.01 \times 10^{-17}$} \\ \hline 
\footnotesize{100} & \footnotesize{256} & \footnotesize{$15 a_0$} & \footnotesize{0.01} & \footnotesize{2.72} & \footnotesize{0.35} & \footnotesize{$1.13 \times 10^{-23}$} & \footnotesize{0.495} & \footnotesize{$5.59 \times 10^{-24}$} \\  
\footnotesize{100} & \footnotesize{256} & \footnotesize{$20 a_0$} & \footnotesize{0.01} & \footnotesize{2.72} & \footnotesize{0.35} & \footnotesize{$1.12 \times 10^{-23}$} & \footnotesize{0.487} & \footnotesize{$5.45 \times 10^{-24}$} \\
\footnotesize{100} & \footnotesize{256} & \footnotesize{$30 a_0$} & \footnotesize{0.01} & \footnotesize{2.72} & \footnotesize{0.35} & \footnotesize{$1.13 \times 10^{-23}$} & \footnotesize{0.493} & \footnotesize{$5.57 \times 10^{-24}$} \\ \hline 
\footnotesize{50} & \footnotesize{512} & \footnotesize{$15 a_0$} & \footnotesize{0.01} & \footnotesize{2.72} & \footnotesize{1} & \footnotesize{$4.16 \times 10^{-27}$} & \footnotesize{0.271} & \footnotesize{$1.13 \times 10^{-27}$} \\
\footnotesize{50} & \footnotesize{512} & \footnotesize{$20 a_0$} & \footnotesize{0.01} & \footnotesize{10.88} & \footnotesize{1} & \footnotesize{$4.35 \times 10^{-27}$} & \footnotesize{0.265} & \footnotesize{$1.15 \times 10^{-27}$} \\
\footnotesize{50} & \footnotesize{512} & \footnotesize{$30 a_0$} & \footnotesize{0.005} & \footnotesize{13.60} & \footnotesize{1} & \footnotesize{$4.52 \times 10^{-27}$} & \footnotesize{0.269} & \footnotesize{$1.22 \times 10^{-27}$} \\ \hline \hline 
\end{tabular} 
\end{center}
\end{table}

\section{Conclusions}

In this paper, on the example of the thermally activated reaction DH + H $\to$ D + H$_2$, we investigated the convergence of the ring polymer molecular dynamics (RPMD) rate constants with respect to several RPMDrate code input parameters at different temperatures (300-50 K). We demonstrated that for the temperatures $T = 100$ and 300 K the standard umbrella integration step $d \xi = 0.01$ and the standard force constant $k_i = 2.72$ ($T/$K) eV, typically used in the RPMDrate code, could be utilized in a wide asymptotic distance range $10a_0 \leq R_{\infty} \leq 30a_0$, the resulting RPMD rate constants are close to each other in all the range of asymptotic distances. As opposed to the temperatures $T = 100$ and 300 K, at the low temperature $T = 50$ K, the umbrella integration step should be decreased and the force constant should be increased (e.g., the modified input RPMDrate code parameters for $R_{\infty} = 30a_0$ are $d \xi = 0.005$ and $k_i = 13.6$ ($T/$K) eV) for obtaining the correct RPMD rate constants if the asymptotic distance is increased. The relative deviation between the resulting RPMD rates obtained for $R_{\infty} = 15a_0$ and $R_{\infty} = 30a_0$ is $\approx 10 \%$ at $T = 50$ K whereas the same deviation is close to zero for $T \geq 100$ K. Thus, if the RPMD rate constant should be computed very accurately then it is necessary to carefully choose the RPMDrate code input parameters, e.g., to increase the asymptotic distance as it was reported for the title reaction. Due to astrochemical importance of the title reaction, lower temperatures (20 K and below) will be the goal of our further study. 
 
\section*{CRediT authorship contribution statement} 

\textbf{Ivan S. Novikov}: Investigation, Visualization, Writing - Original draft, Data Curation. \textbf{Yury V. Suleimanov}: Conceptualization, Methodology, Software, Writing - Review \& Editing. \textbf{Alexander V. Shapeev}: Conceptualization, Resources, Writing - Review \& Editing, Project administration, Funding acquisition, Supervision. 
 
\section*{Declaration of Competing Interest}

None.
 
\section*{Acknowledgements}

The work of the authors was supported by the Russian Foundation for Basic Research (grant number 20-03-00833). Y.V.S. also acknowledges the European Regional Development Fund and the Republic of Cyprus through the Research Promotion Foundation (Projects: INFRASTRUCTURE/1216/0070 and Cy-Tera NEAY$\Pi$O$\Delta$OMH/$\Sigma$TPATH/0308/31). 

\bibliographystyle{plain}
\bibliography{article}

\begin{thebibliography}{10}

\bibitem{allen2013communication}
Joshua~W Allen, William~H Green, Yongle Li, Hua Guo, and Yury~V Suleimanov.
\newblock Communication: Full dimensional quantum rate coefficients and kinetic
  isotope effects from ring polymer molecular dynamics for a seven-atom
  reaction {OH+CH$_4$ $\rightarrow $ CH$_3$+H$_2$O}, 2013.

\bibitem{andersen1983rattle}
Hans~C Andersen.
\newblock Rattle: A “velocity” version of the shake algorithm for molecular
  dynamics calculations.
\newblock {\em Journal of Computational Physics}, 52(1):24--34, 1983.

\bibitem{aoiz2006cumulative}
FJ~Aoiz, M~Brouard, CJ~Eyles, JF~Castillo, and V~S{\'a}ez~R{\'a}banos.
\newblock Cumulative reaction probabilities: A comparison between
  quasiclassical and quantum mechanical results.
\newblock {\em The Journal of chemical physics}, 125(14):144105, 2006.

\bibitem{bennett1977molecular}
Charles~H Bennett.
\newblock Molecular dynamics and transition state theory: the simulation of
  infrequent events.
\newblock ACS Publications, 1977.

\bibitem{bhowmick2018}
Somnath Bhowmick, Duncan Bossion, Yohann Scribano, and Yury~V Suleimanov.
\newblock The low temperature {D$^+$+H$_2$ $\rightarrow $ HD+H$^+$} reaction
  rate coefficient: A ring polymer molecular dynamics and quasi-classical
  trajectory study.
\newblock {\em Physical Chemistry Chemical Physics}, 20(41):26752--26763, 2018.

\bibitem{boothroyd1996h3}
Arnold~I Boothroyd, William~J Keogh, Peter~G Martin, and Michael~R Peterson.
\newblock A refined {H}$_3$ potential energy surface.
\newblock {\em The Journal of chemical physics}, 104(18):7139--7152, 1996.

\bibitem{bowman2001importance}
Joel~M Bowman, Dunyou Wang, Xinchuan Huang, Fermin Huarte-Larra{\~n}aga, and
  Uwe Manthe.
\newblock The importance of an accurate ch 4 vibrational partition function in
  full dimensionality calculations of the {H+CH$_4$ $\rightarrow$ H$_2$+CH$_3$}
  reaction.
\newblock {\em The Journal of Chemical Physics}, 114(21):9683--9684, 2001.

\bibitem{Suleimanov2017A}
JF~Castillo and YV~Suleimanov.
\newblock A ring polymer molecular dynamics study of the {OH+H{$_2$}(D{$_2$})}
  reaction.
\newblock {\em Phys. Chem. Chem. Phys.}, 19(43):29170--29176, 2017.

\bibitem{ceriotti2009langevin}
Michele Ceriotti, Giovanni Bussi, and Michele Parrinello.
\newblock Langevin equation with colored noise for constant-temperature
  molecular dynamics simulations.
\newblock {\em Physical review letters}, 102(2):020601, 2009.

\bibitem{ceriotti2009colored}
Michele Ceriotti, Giovanni Bussi, and Michele Parrinello.
\newblock Nuclear quantum effects in solids using a colored-noise thermostat.
\newblock {\em Physical review letters}, 103(3):030603, 2009.

\bibitem{ceriotti2010colored}
Michele Ceriotti, Giovanni Bussi, and Michele Parrinello.
\newblock Colored-noise thermostats {\`a} la carte.
\newblock {\em Journal of Chemical Theory and Computation}, 6(4):1170--1180,
  2010.

\bibitem{chandler1978}
David Chandler.
\newblock Statistical mechanics of isomerization dynamics in liquids and the
  transition state approximation.
\newblock {\em J. Chem. Phys.}, 68(6):2959--2970, 1978.

\bibitem{Wolynes}
David Chandler and Peter~G Wolynes.
\newblock Exploiting the isomorphism between quantum theory and classical
  statistical mechanics of polyatomic fluids.
\newblock {\em J. Chem. Phys.}, 74(7):4078--4095, 1981.

\bibitem{Manolopoulos2004}
Ian~R Craig and David~E Manolopoulos.
\newblock Quantum statistics and classical mechanics: Real time correlation
  functions from ring polymer molecular dynamics.
\newblock {\em J. Chem. Phys.}, 121(8):3368--3373, 2004.

\bibitem{Manolopoulos2005_122}
Ian~R Craig and David~E Manolopoulos.
\newblock Chemical reaction rates from ring polymer molecular dynamics.
\newblock {\em The Journal of chemical physics}, 122(8):084106, 2005.

\bibitem{Manolopoulos2005_123}
Ian~R Craig and David~E Manolopoulos.
\newblock A refined ring polymer molecular dynamics theory of chemical reaction
  rates.
\newblock {\em J. Chem. Phys.}, 123(3):034102, 2005.

\bibitem{dalgarno1972infrared}
A~Dalgarno and EL~Wright.
\newblock Infrared emissivities of {H}$_2$ and {HD}.
\newblock {\em The Astrophysical Journal}, 174:L49, 1972.

\bibitem{espinosa2020vtst}
Joaquin Espinosa-Garcia, Moises Garcia-Chamorro, Jose~Carlos Corchado, Somnath
  Bhowmick, and Yury Suleimanov.
\newblock Vtst and rpmd kinetics study of the nine-body {X+C$_2$H$_6$
  (X=H,Cl,F)} reactions based on analytical potential energy surfaces.
\newblock {\em Physical Chemistry Chemical Physics}, 2020.

\bibitem{gillan87b}
M~J Gillan.
\newblock Quantum-classical crossover of the transition rate in the damped
  double well.
\newblock {\em J. Phys. C}, 20:3621--3641, August 1987.

\bibitem{gillan87a}
M.~J. Gillan.
\newblock Quantum simulation of hydrogen in metals.
\newblock {\em Phys. Rev. Lett.}, 58(6):563--566, February 1987.

\bibitem{gonzalez2014theoretical}
Eloisa Gonzalez-Lavado, Jose~C Corchado, Yury~V Suleimanov, William~H Green,
  and Joaquin Espinosa-Garcia.
\newblock Theoretical kinetics study of the {O($^{3}$P)+CH$_4$/CD$_4$} hydrogen
  abstraction reaction: the role of anharmonicity, recrossing effects, and
  quantum mechanical tunneling.
\newblock {\em The Journal of Physical Chemistry A}, 118(18):3243--3252, 2014.

\bibitem{RPMDREVIEW}
Scott Habershon, David~E. Manolopoulos, Thomas~E. Markland, and Thomas
  F.~Miller III.
\newblock Ring-polymer molecular dynamics: Quantum effects in chemical dynamics
  from classical trajectories in an extended phase space.
\newblock {\em Annu. Rev. Phys. Chem.}, 64(1):387--413, 2013.

\bibitem{hazra2016gp_effect_h2d}
Jisha Hazra, Brian~K Kendrick, and Naduvalath Balakrishnan.
\newblock Geometric phase effects in ultracold hydrogen exchange reaction.
\newblock {\em Journal of Physics B: Atomic, Molecular and Optical Physics},
  49(19):194004, 2016.

\bibitem{hickson2015ch2}
Kevin~M. Hickson, Jean-Christophe Loison, Hua Guo, and Yury~V. Suleimanov.
\newblock Ring-polymer molecular dynamics for the prediction of low-temperature
  rates: An investigation of the {C($^1$D)+H$_2$} reaction.
\newblock {\em J. Phys. Chem. Lett.}, 6(21):4194--4199, 2015.

\bibitem{hickson2017experimentalcd2}
Kevin~M Hickson and Yury~V Suleimanov.
\newblock An experimental and theoretical investigation of the {C($^1$D) +
  D$_2$} reaction.
\newblock {\em Physical Chemistry Chemical Physics}, 19(1):480--486, 2017.

\bibitem{hickson2017oh2}
Kevin~M Hickson and Yury~V Suleimanov.
\newblock Low-temperature experimental and theoretical rate constants for the
  {O ($^1$D)+ H$_2$} reaction.
\newblock {\em The Journal of Physical Chemistry A}, 121(9):1916--1923, 2017.

\bibitem{Kastner2009}
Johannes K{\"a}stner.
\newblock Umbrella integration in two or more reaction coordinates.
\newblock {\em J. Chem. Phys.}, 131(3):034109, 2009.

\bibitem{Kastner2005}
Johannes K{\"a}stner and Walter Thiel.
\newblock Bridging the gap between thermodynamic integration and umbrella
  sampling provides a novel analysis method:“umbrella integration”.
\newblock {\em J. Chem. Phys.}, 123(14):144104, 2005.

\bibitem{Kastner2006}
Johannes K{\"a}stner and Walter Thiel.
\newblock Analysis of the statistical error in umbrella sampling simulations by
  umbrella integration.
\newblock {\em J. Chem. Phys.}, 124(23):234106, 2006.

\bibitem{lepp2002atomic}
Stephen Lepp, PC~Stancil, and A~Dalgarno.
\newblock Atomic and molecular processes in the early universe.
\newblock {\em Journal of Physics B: Atomic, Molecular and Optical Physics},
  35(10):R57, 2002.

\bibitem{yongle2014quantum}
Yongle Li, Yury~V Suleimanov, William~H Green, and Hua Guo.
\newblock Quantum rate coefficients and kinetic isotope effect for the reaction
  {Cl+CH$_4$ $\rightarrow $ HCl+CH$_3$} from ring polymer molecular dynamics.
\newblock {\em The Journal of Physical Chemistry A}, 118(11):1989--1996, 2014.

\bibitem{mcgreer2008impact}
Ian~D McGreer and Greg~L Bryan.
\newblock The impact of {HD} cooling on the formation of the first stars.
\newblock {\em The Astrophysical Journal}, 685(1):8, 2008.

\bibitem{miller1983}
William~H Miller, Steven~D Schwartz, and John~W Tromp.
\newblock Quantum mechanical rate constants for bimolecular reactions.
\newblock {\em The Journal of chemical physics}, 79(10):4889--4898, 1983.

\bibitem{RPMD:MuH}
Ricardo P\'erez~de Tudela, F.~J. Aoiz, Yury~V. Suleimanov, and David~E.
  Manolopoulos.
\newblock Chemical reaction rates from ring polymer molecular dynamics: Zero
  point energy conservation in {Mu + H$_2$ $\rightarrow $ MuH + H}.
\newblock {\em J. Phys. Chem. Let.}, 3(4):493--497, 2012.

\bibitem{RPMD:DMuH}
Ricardo P\'erez~de Tudela, Yury~V. Suleimanov, Jeremy~O. Richardson, Vicente
  S\'aez~R\'abanos, William~H. Green, and F.~J. Aoiz.
\newblock Stress test for quantum dynamics approximations: Deep tunneling in
  the muonium exchange reaction {D+HMu $\rightarrow $ DMu+H}.
\newblock {\em J. Phys. Chem. Lett.}, 5(23):4219--4224, 2014.

\bibitem{rampino2016thermal}
Sergio Rampino and Yury~V Suleimanov.
\newblock Thermal rate coefficients for the astrochemical process {C+CH$^+$
  $\rightarrow $ C$_2^+$+H} by ring polymer molecular dynamics.
\newblock {\em The Journal of Physical Chemistry A}, 120(50):9887--9893, 2016.

\bibitem{Manolopoulos2011}
Yury~V Suleimanov, Rosana Collepardo-Guevara, and David~E Manolopoulos.
\newblock Bimolecular reaction rates from ring polymer molecular dynamics:
  Application to {{H}+CH{$_4$}} $\to$ {H{$_2$}+CH{$_3$}}.
\newblock {\em J. Chem. Phys.}, 134(4):044131, 2011.

\bibitem{suleimanov2014XH2}
Yury~V Suleimanov, Wendi~J Kong, Hua Guo, and William~H Green.
\newblock Ring-polymer molecular dynamics: Rate coefficient calculations for
  energetically symmetric (near thermoneutral) insertion reactions {(X+H$_2$)
  $\rightarrow $ HX+H (X=C($^1$D),S($^1$D))}.
\newblock {\em The Journal of chemical physics}, 141(24):244103, 2014.

\bibitem{RPMDrate}
Yu.V. Suleimanov, J.W. Allen, and W.H. Green.
\newblock {RPMDrate}: Bimolecular chemical reaction rates from ring polymer
  molecular dynamics.
\newblock {\em Comp. Phys. Comm.}, 184(3):833--840, 2013.

\bibitem{uehara2000dwarf}
Hideya Uehara and Shu-ichiro Inutsuka.
\newblock Does deuterium enable the formation of primordial brown dwarfs?
\newblock {\em The Astrophysical Journal Letters}, 531(2):L91, 2000.

\bibitem{voth89}
Gregory~A. Voth, David Chandler, and William~H. Miller.
\newblock Rigorous formulation of quantum transition state theory and its
  dynamical corrections.
\newblock {\em J. Chem. Phys.}, 91:7749, 1989.

\bibitem{yuan2020gp_effect_h2d}
Daofu Yuan, Yin Huang, Wentao Chen, Hailin Zhao, Shengrui Yu, Chang Luo, Yuxin
  Tan, Siwen Wang, Xingan Wang, Zhigang Sun, et~al.
\newblock Observation of the geometric phase effect in the {H+ HD} $\rightarrow
  $ {H}$_2$+ {D} reaction below the conical intersection.
\newblock {\em Nature communications}, 11(1):1--7, 2020.

\bibitem{zhou2020feshbach_resonance_h2d}
Boyi Zhou, Benhui Yang, Naduvalath Balakrishnan, Brian~K Kendrick, and
  Phillip~C Stancil.
\newblock Prediction of a feshbach resonance in the below-the-barrier reactive
  scattering of vibrational excited {HD} with {H}.
\newblock {\em The Journal of Physical Chemistry Letters}, 2020.

\end{thebibliography}

\end{document}